\documentclass[aps,prl,reprint,superscriptaddress,nofootinbib,floatfix]{revtex4-2}
\usepackage{amsmath,amssymb,bm}
\usepackage{graphicx}
\usepackage{tikz}
\usetikzlibrary{arrows.meta,positioning,calc}
\usepackage{microtype}
\newcommand{\p}{\mathbf p}
\newcommand{\pperp}{\mathbf p_\perp}
\newcommand{\sig}{\boldsymbol\sigma}
\newcommand{\al}{\boldsymbol\alpha}

\begin{document}

\title{Spectral Reduction Selects the Observable Imprint of Dirac Quantum Geometry}

\author{Zhongze Guo}
\email{guozhongze007@gmail.com}
\affiliation{Department of Physics and Institute of Theoretical Physics, University of Science and Technology Beijing, Beijing 100083, China}

\author{Bei Xu}
\email{xubei0903@163.com}
\affiliation{Institute for Advanced Study, Tsinghua University, Beijing 100084, China}	

\author{Qiang Gu}
\email[Corresponding author: ] {qgu@ustb.edu.cn}
\affiliation{Department of Physics and Institute of Theoretical Physics, University of Science and Technology Beijing, Beijing 100083, China}

\begin{abstract}
Spectral constraints can alter the observable manifestation of quantum geometry without changing the local eigenspace. Relativistic electron vortices provide a sharp example: the spin orbit interaction redistributes spin and orbital angular momentum, yet its magnetic contribution differs between localized three-dimensional wave packets and stationary monochromatic beams. We resolve this discrepancy by showing that fixing the energy replaces ambient momentum derivatives in position-weighted current moments by shell-tangent derivatives and derive a Hermitian selection rule for first-order observables in momentum-dependent internal subspaces. Its extension-independent form applies beyond Dirac systems, including a broad class of antisymmetric first moments of scalar-charge velocity currents. For the free positive-energy Dirac doublet, the magnetic correction is an exact non-Abelian Berry-curvature contraction with a pointwise saturable quantum-metric bound. Changing the relative phase between the two $\sigma_z$ spinor components switches this correction from zero in a conserved transverse-helicity sector to a finite value in a radial coherent superposition. At fixed finite Lorentz factor, the radial response scales as $O(\theta/\gamma)$, compared with $O(\theta^2)$ for longitudinal polarization. Spectral reduction and internal coherence thus provide distinct controls of the observable imprint of quantum geometry.
\end{abstract}

\maketitle

\paragraph{Introduction.—}

Momentum-dependent internal eigenspaces underlie Berry transport,
non-Abelian adiabatic structure, and quantum-metric response in multicomponent
quantum systems~\cite{Berry1984,WilczekZee1984,ProvostVallee1980,Xiao2010,Kolodrubetz2017,Gao2026}.
For positive-energy Dirac electrons, the geometric structure of spin--orbit
interaction (SOI) has been extensively studied in wave-packet dynamics and vortex
states~\cite{Mathur1991,Bliokh2005,Chuu2010,Bliokh2011,BliokhPhysRep2017,Lloyd2017}.
Yet the quantum geometry of an eigenspace does not by itself determine how it
appears in a particular observable. This issue becomes especially sharp under
spectral restriction, which changes the momentum derivatives entering
position-weighted observables while leaving the local eigenspace unchanged.
Relativistic electron vortices provide a clean setting for this question and
have been realized experimentally
~\cite{Bliokh2007,Bliokh2011,BliokhPRA2017,Uchida2010,Verbeeck2010,McMorran2011}.

For a free positive-energy Dirac vortex with azimuthal index $\ell$ and
rest-frame spin projection $s=\pm\tfrac12$ along the vortex axis, the
single-cone expectation values are
$S_z=\hbar s(1-\Delta)$ and $L_z=\hbar(\ell+s\Delta)$, with
$\Delta=c^2p_\perp^2/[E(E+mc^2)]$.
The SOI transfers angular momentum $\hbar s\Delta$
from spin to orbital motion while leaving $J_z=\hbar(\ell+s)$ unchanged.
The magnetic moment, however, exposes a spectral distinction:
\begin{equation}
	M_z^{\rm amb}=\mu_E(\ell+2s-s\Delta),\qquad
	M_z^{(E)}=\mu_E(\ell+2s),
	\label{eq:magneticcontrast}
\end{equation}
where $\mu_E=q\hbar c^2/(2E)$.
Their difference, $-\mu_Es\Delta$, is the $O(\Delta)$ magnetic-moment
discrepancy between one-particle expectation values and normalized
monochromatic-beam linear densities highlighted in
Ref.~\cite{Bliokh2011}.

This packet--beam mismatch points to a broader mechanism. Fixing the energy does not alter the local positive-energy projector, but changes the momentum-space direction along which a position-weighted observable differentiates the state. More generally, for first-order observables whose normal intraband symbol vanishes in a neighborhood of the energy shell, we derive an extension-independent Hermitian correction controlled by the observable's inter-subspace vertex and the projector variation along the eliminated momentum direction. For the free Dirac magnetic moment, the correction is an exact non-Abelian Berry-curvature contraction, while the quantum metric sets a pointwise saturable bound on its magnitude.

The same mechanism predicts a symmetry-resolved response. The shell-induced
magnetic correction anticommutes with the conserved transverse-helicity
operator, so each resolved branch is exactly dark to this correction, whereas
a radial coherent superposition of the two branches is bright. For small cone angle $\theta$ at fixed finite Lorentz factor
$\gamma=E/(mc^2)$, the radial response scales as $O(\theta/\gamma)$,
compared with $O(\theta^2)$ for longitudinal polarization. Spectral reduction can therefore
leave the local eigenspace unchanged while altering how its quantum geometry
enters a position-weighted current moment.

\paragraph{Geometric structure of the Dirac magnetic moment.—}
To identify what spectral reduction can change, we first separate the
vortex-phase contribution from the spinor-derivative contribution. We use the
following orthonormal frame for the twofold-degenerate positive-energy
eigenspace,
\begin{equation}
	U(\p)=\sqrt{\frac{E+mc^2}{2E}}
	\begin{pmatrix}
		\mathbf{1}_2\\[2mm]
		\dfrac{c\sig\cdot\p}{E+mc^2}
	\end{pmatrix},
	\qquad
	U^\dagger\al U=\frac{c\p}{E}\mathbf{1}_2 .
	\label{eq:U}
\end{equation}
The ambient transverse first moment of the Dirac current is represented by
\begin{equation}
	\hat M_z^{\rm amb}=\frac{qi\hbar c}{2}
	(\alpha_y\partial_{p_x}-\alpha_x\partial_{p_y}).
	\label{eq:Mambientop}
\end{equation}
Acting on states of the form
$e^{i\ell\phi_p}f(p_\perp,p_z)U(\p)\eta$ and projecting onto the
positive-energy doublet, the radial-envelope derivatives vanish because
$U^\dagger(p_x\alpha_y-p_y\alpha_x)U=0$. One obtains
\begin{equation}
	\mathsf M_z^{\rm amb}
	=\mu_E\ell\,\mathbf{1}_2+\mu_E\sigma_z+\mathcal G_z(\p),
	\label{eq:ambientkernel}
\end{equation}
where the complete spinor-derivative contribution is
\begin{align}
	&\frac{qi\hbar c}{2}
	U^\dagger(\alpha_y\partial_{p_x}-\alpha_x\partial_{p_y})U
	=\mu_E\sigma_z+\mathcal G_z(\p),
	\nonumber\\[-1mm]
	&\mathcal G_z(\p)
	=-\frac{q\hbar c^4}{4E^2(E+mc^2)}
	\left[p_\perp^2\sigma_z-p_z\pperp\!\cdot\!\sig_\perp\right].
	\label{eq:Ggeom}
\end{align}
Thus $\mathcal G_z$ is the residual geometry-dependent kernel entering the
magnetic observable after the usual $\mu_E\sigma_z$ spin term has been
separated. For longitudinal polarization,
$\sigma_z\eta_s=2s\eta_s$,
\begin{equation*}
	\eta_s^\dagger\mathcal G_z\eta_s=-\mu_Es\Delta,
\end{equation*}
which recovers the ambient result in Eq.~\eqref{eq:magneticcontrast}.
The familiar $s\Delta$ contribution is therefore only one
polarization-resolved matrix element of the full magnetic kernel.

Exact restriction to $E=E_0$ leaves the local positive-energy subspace
unchanged but replaces the ambient momentum derivatives by derivatives tangent
to the fixed-energy shell. The next question is how this replacement changes
the projected observable itself.

\paragraph{Observable selection by spectral reduction.—}
Let $P(\p)=U(\p)U^\dagger(\p)$ be an isolated internal subspace with scalar
dispersion $\varepsilon(\p)$, restricted to
$\Sigma_E:\varepsilon(\p)=E_0$. In a regular spectral chart with
$v_n\equiv\partial_{p_n}\varepsilon\neq0$, solve locally for $p_n$ and write
$v_i=\partial_{p_i}\varepsilon$. Derivatives tangent to the fixed-energy shell
are
\begin{equation}
	D_a\equiv\left.\partial_{p_a}\right|_E
	=\partial_{p_a}-\frac{v_a}{v_n}\partial_{p_n}.
	\label{eq:tangentgeneral}
\end{equation}
Here ``normal'' denotes the eliminated coordinate direction of this local
spectral chart, not necessarily the orthogonal Euclidean normal to the shell.

Consider a symmetrically ordered first-order observable
\begin{equation}
	\hat O_3=W+i\hbar\sum_a
	\left(B_a\partial_{p_a}+\frac12\partial_{p_a}B_a\right),
	\label{eq:Oambient}
\end{equation}
with $W=W^\dagger$ and $B_a=B_a^\dagger$, where $a$ labels coordinates
retained on the shell. Define
\begin{equation}
	B_N=\sum_a\frac{v_a}{v_n}B_a,
	\qquad Q=1-P.
	\label{eq:BN}
\end{equation}
The shell restriction removes the derivative term
$i\hbar B_N\partial_{p_n}$. For a state $\Psi=U\eta$,
\begin{equation}
	\partial_{p_n}(U\eta)=
	U\bigl(\partial_{p_n}+U^\dagger\partial_{p_n}U\bigr)\eta
	+Q(\partial_{p_n}U)\eta .
	\label{eq:normaldecomp}
\end{equation}
The first term is intraband and can depend on how the shell state is continued
away from the shell, whereas the second measures the inter-subspace variation
of the projector along the eliminated spectral direction.

Under the local coarea transform,
$d^dp=dE\,d^{d-1}p_a/|v_n|$, followed by the corresponding half-density
rescaling,
\[
\begin{aligned}
	\mathcal U_c(\hat O_3-W)\mathcal U_c^{-1}
	&=i\hbar\sum_a\left(B_aD_a+\frac12D_aB_a\right)\\
	&\quad+i\hbar\left(B_E\partial_E+\frac12\partial_EB_E\right),\\
	B_E&\equiv\sum_a v_aB_a=v_nB_N .
\end{aligned}
\]
Spectral reduction is the exact passage to one fixed-energy fiber while
retaining the shell-tangent Hermitian operator. If the eliminated derivative
has no intraband matrix element in a neighborhood of the shell,
$PB_NP=0$, the envelope derivative drops out. Differentiating
$PB_EP=0$ then combines with the symmetric-ordering term to give
\begin{equation}
	\delta O_{\rm geom}
	=-\frac{i\hbar}{2}
	\left(\mathcal C_N-\mathcal C_N^\dagger\right),
	\qquad
	\mathcal C_N=PB_NQ(\partial_{p_n}P)P .
	\label{eq:selectiontheorem}
\end{equation}
Here $PB_NQ$ is the inter-subspace coupling selected by the observable,
whereas $Q(\partial_{p_n}P)P$ measures the projector variation along the
eliminated spectral direction. Both must be present, but only the
anti-Hermitian part of their product contributes. If $PB_NP\neq0$, additional
terms can depend on how the shell state is continued away from the shell.
Equation~\eqref{eq:selectiontheorem} is frame independent because it is
written entirely in terms of $P$, $Q$, and the ambient observable.

The same inter-subspace derivatives define the non-Abelian quantum geometric
tensor~\cite{ProvostVallee1980,WilczekZee1984,Ma2010,Kolodrubetz2017,PeottaTorma2015},
\begin{equation}
	\mathsf T_{ij}=(Q\partial_iU)^\dagger(Q\partial_jU)
	=\mathsf g_{ij}-\frac{i}{2}\mathsf F_{ij}.
	\label{eq:QGT}
\end{equation}
Writing
$\delta\mathfrak O_{\rm geom}=U^\dagger\delta O_{\rm geom}U$,
for a normalized internal spinor $\chi$ one finds
\begin{equation}
	|\chi^\dagger\delta\mathfrak O_{\rm geom}\chi|^2
	\le \hbar^2
	(\chi^\dagger\mathsf V_N\chi)
	(\chi^\dagger\mathsf g_{nn}\chi),
	\label{eq:metricbound}
\end{equation}
where $\mathsf V_N=U^\dagger B_NQB_NU$. Thus, for a fixed observable
vertex, the quantum metric bounds the accessible geometric response. For the
free Dirac magnetic moment this bound is pointwise saturable.

Importantly, the condition $PB_NP=0$ is not specific to Dirac systems.
For a smooth isolated band or exactly degenerate subspace with canonical
position and scalar-charge current $q\,\partial_{\p}H$, the antisymmetric
first moment of the current satisfies
$PB_EP=(q/2)(v_xv_y-v_yv_x)P=0$ identically. The same reduction rule therefore
applies to this broader class of current moments. In general the correction
contains an interband-gap-weighted geometric tensor; the free Dirac case is
special because its common complementary gap reduces this tensor to the
non-Abelian quantum geometric tensor. See Supplemental Material~\cite{SupplementalMaterial} for the coarea
derivation, velocity-current generalization, controlled packet-to-shell
comparison, regularity conditions, shell pullback, fixed-$p_z$ control, and
stationary Gordon-current check.

\begin{figure}[t]
\centering
\includegraphics[width=0.98\columnwidth]{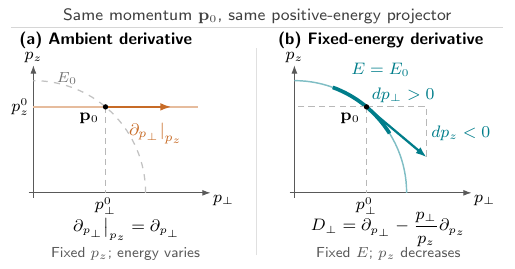}
\caption{Exact spectral reduction changes the derivative entering a
position-weighted observable, not the local positive-energy frame. Panel (a)
uses the ambient derivative at fixed $p_z$, whereas panel (b) restricts to the
fixed-energy shell and uses the shell-tangent derivative; the tangent lies along the forward
constant-energy arc and points toward decreasing $p_z$ as $p_\perp$
increases.}
\label{fig:spectral}
\end{figure}

\paragraph{Dirac magnetic moment: dark and bright sectors.—}
On the regular forward-propagating patch $p_z>0$, the fixed-energy derivatives
are
$D_a=\partial_{p_a}-(p_a/p_z)\partial_{p_z}$, while the associated normal
intraband symbol satisfies $PB_NP=0$ throughout the regular patch by
Eq.~\eqref{eq:U}. The shell-induced correction is therefore extension
independent and is given by Eq.~\eqref{eq:selectiontheorem}. Direct evaluation
gives
\begin{equation}
	\delta\mathsf M_z
	=\frac{q\hbar c^2}{4Ep_z}\,\pperp\!\cdot\!\sig_\perp
	-\mathcal G_z(\p).
	\label{eq:shellcorr}
\end{equation}
Equivalently, in terms of the non-Abelian curvature defined by
Eq.~\eqref{eq:QGT},
\begin{equation}
	\delta\mathsf M_z
	=-\frac{q\hbar E}{2p_z}
	\left(p_x\mathsf F_{yz}-p_y\mathsf F_{xz}\right).
	\label{eq:curvatureM}
\end{equation}
Thus the current vertex selects a specific curvature contraction.
Equation~\eqref{eq:shellcorr} also makes the cancellation mechanism explicit:
when added to the ambient kernel, the shell correction contains
$-\mathcal G_z$, canceling the residual geometry-dependent part of the
spinor derivative in Eq.~\eqref{eq:Ggeom}, while the usual
$\mu_E\sigma_z$ spin term remains and a distinct transverse-spin term is
generated. For longitudinal polarization the latter has zero expectation
value, giving
\begin{equation}
	M_z^{(E)}=\mu_E(\ell+2s),\qquad
	M_z^{(E)}-M_z^{\rm amb}=+\mu_Es\Delta .
	\label{eq:ME}
\end{equation}
The $s\Delta$ spin--OAM redistribution itself is unchanged; spectral
reduction removes only its contribution to this magnetic observable.
A transversely normalizable fixed-$p_z$ state instead retains the ambient
derivatives and hence the transverse-momentum average of the ambient result
in Eq.~\eqref{eq:magneticcontrast}. The packet--beam distinction is therefore
spectral rather than a consequence of longitudinal delocalization
(Fig.~\ref{fig:spectral}).

The full matrix structure of Eq.~\eqref{eq:shellcorr} reveals a stronger
selection rule. Write $p_\perp=p\sin\theta$, $p_z=p\cos\theta$, and
$\gamma=E/(mc^2)$. In the cylindrical Pauli basis,
\begin{equation}
	\delta\mathsf M_z=A_r\sigma_r+A_z\sigma_z,
	\label{eq:MrMz}
\end{equation}
with
\begin{align}
	A_r&=\frac{\mu_E}{2}\tan\theta
	\left(\sin^2\theta+\frac{\cos^2\theta}{\gamma}\right),\nonumber\\
	A_z&=\frac{\mu_E}{2}\left(1-\frac1\gamma\right)\sin^2\theta .
	\label{eq:ArAz}
\end{align}
Introduce the conserved transverse-helicity operator
$\hat K=\beta(\bm\Sigma\times\p)_z$, used in cylindrical Dirac mode
constructions~\cite{Khosravi2019}. On the positive-energy doublet,
\begin{equation}
	\mathsf K\equiv
	U^\dagger\frac{\hat K}{p_\perp}U=-\sigma_\phi,
	\qquad p_\perp\neq0 .
	\label{eq:Kproj}
\end{equation}
Since $\sigma_\phi$ anticommutes with both $\sigma_r$ and $\sigma_z$,
\begin{equation}
	\{\delta\mathsf M_z,\mathsf K\}=0,\qquad
	\Pi_\pm\delta\mathsf M_z\Pi_\pm=0,
	\label{eq:Kdark}
\end{equation}
where $\Pi_\pm=(1\pm\mathsf K)/2$. Every resolved transverse-helicity branch
is therefore exactly dark to the shell-induced correction, not merely
suppressed. Coherence between opposite branches is necessary, though not
sufficient, for a nonzero response. Here ``dark'' refers to
$\delta\mathsf M_z$, not to the total magnetic moment.

\begin{figure*}[!t]
	\centering
	\includegraphics[width=0.94\textwidth]{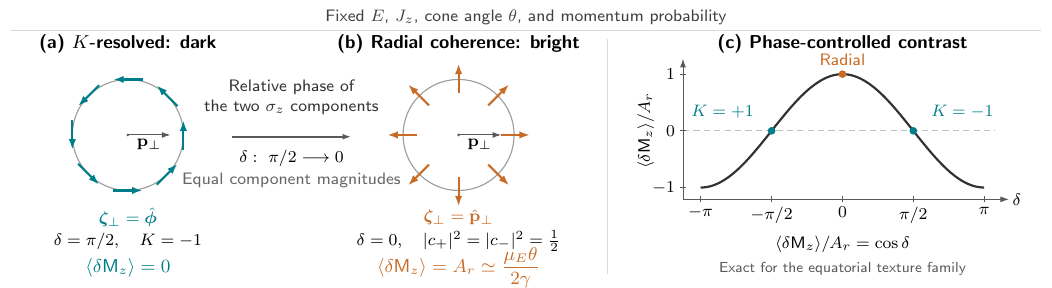}
	\caption{At fixed $E$, $J_z$, cone angle, and momentum distribution, resolved
		transverse-helicity branches are exactly dark,
		$\Pi_\pm\delta\mathsf M_z\Pi_\pm=0$. Varying the relative phase of the two
		$\sigma_z$ components continuously controls the response, with
		$\langle\delta M_z\rangle/A_r=\cos\delta$ for the texture family in
		Eq.~\eqref{eq:texturefamily}; the radial coherent state corresponds to
		$\delta=0$.}
	\label{fig:Kdarkbright}
\end{figure*}

This structure yields a controlled bright--dark pair at fixed spectrum and
fixed $J_z$. Consider
\begin{equation}
	\eta_{\ell,\delta}(\phi_p)
	=\frac{e^{i\ell\phi_p}}{\sqrt2}
	\begin{pmatrix}
		1\\ e^{i(\phi_p+\delta)}
	\end{pmatrix},
	\qquad
	J_z=\hbar\left(\ell+\frac12\right).
	\label{eq:texturefamily}
\end{equation}
Its transverse polarization is
$\boldsymbol\zeta_\perp
=\cos\delta\,\hat{\mathbf p}_\perp
+\sin\delta\,\hat{\boldsymbol\phi}$, and hence
\begin{equation}
	\delta M_z(\delta)=A_r\cos\delta .
	\label{eq:deltafamily}
\end{equation}
The azimuthal states $\delta=\pm\pi/2$ are the two definite-$K$ branches and
are exactly dark, whereas the radial state $\delta=0$, an equal-weight
coherent superposition of them, is bright:
\begin{equation}
	\delta M_z^{\rm rad}=A_r
	=\frac{\mu_E}{2\gamma}\theta+O(\theta^3),
	\label{eq:radial}
\end{equation}
at fixed finite $\gamma$. By comparison, longitudinal polarization probes
$A_z$ and gives
$\delta M_z=\mu_Es\Delta=O(\theta^2)$.
The familiar $s\Delta$ response is therefore a subleading polarization
channel within a broader coherence-controlled response. At fixed $E$, $J_z$,
cone angle, and momentum distribution, changing only the relative phase
switches the differential packet--beam response from zero in a resolved-$K$
state to
$\delta M_z^{\rm rad}\simeq\mu_E\theta/(2\gamma)$ in the radial coherent state.

\paragraph{Discussion.—}
Equation~\eqref{eq:selectiontheorem} separates three structures that are
easily conflated: the geometry of the eigenspace, the spectral constraint,
and the observable used to probe it. Spectral reduction leaves the local
projector unchanged and pulls its geometry back to the shell,
$\mathsf g^\Sigma_{ab}=t_a^{\,i}t_b^{\,j}\mathsf g_{ij}$, but changes the
derivative entering a position-dependent observable. Nonzero quantum geometry
alone therefore does not guarantee a nonzero observable response; the
observable vertex and the Hermitian combination selected by
Eq.~\eqref{eq:selectiontheorem} determine what survives.

The free Dirac magnetic moment provides an exact geometric closure.
Equation~\eqref{eq:curvatureM} expresses the signed response as a
non-Abelian Berry-curvature contraction, while
Eq.~\eqref{eq:metricbound} bounds its magnitude by the quantum metric; the
bound is pointwise saturable by an appropriate internal polarization. For the
positive-energy doublet,
\[
\mathsf g_{\varphi\varphi}=g_{\varphi\varphi}\mathbf{1}_2,
\qquad
g_{\varphi\varphi}=\frac{c^2p_\perp^2}{4E^2},
\]
and hence
\begin{equation}
	\Delta=\frac{4E}{E+mc^2}\,g_{\varphi\varphi}.
	\label{eq:Deltametric}
\end{equation}
Thus the familiar Dirac spin--orbit scale $\Delta$ itself has an exact
angular quantum-metric interpretation, although its magnetic manifestation
depends on both polarization and spectral reduction.

The symmetry-resolved response makes this selectivity explicit. Definite
transverse-helicity sectors are exactly dark to the shell-induced correction
despite nontrivial quantum geometry, whereas a radial coherent superposition
is bright. The longitudinal $s\Delta=O(\theta^2)$ response is therefore only
one polarization-resolved component of the full matrix correction, whose
radial coherent component begins at $O(\theta/\gamma)$ for fixed finite
$\gamma$.

An independent coordinate-space Gordon-current calculation yields the same
$M_z^{(E)}=\mu_E(\ell+2s)$, confirming that the cancellation is not an
artifact of the momentum-space representation. The mechanism is also not
specific to the Dirac algebra: for antisymmetric first moments of a
scalar-charge velocity current $q\,\partial_{\mathbf p}H$, the key condition
$PB_EP=0$ follows identically for an isolated scalar-dispersion subspace.
More generally the correction involves an interband-gap-weighted geometric
tensor; the free Dirac problem is special because its common complementary
gap reduces this structure exactly to the non-Abelian quantum geometric
tensor. Exact Dirac--Landau vortices provide a complementary field-dressed
setting in which the integrated magnetic moment is governed by the
gauge-covariant $L_z+2S_z$ combination~\cite{vanKruining2017}.

The broader lesson is that a spectral constraint can change a
position-dependent observable without changing the underlying local
eigenspace. In relativistic electron vortices, this principle explains the
packet--beam magnetic-moment discrepancy and predicts exact dark sectors
together with a phase-controlled radial response. Spectral reduction and
internal coherence therefore provide distinct controls of how quantum
geometry enters an observable.

\textit{Acknowledgment}
Q.G. acknowledges financial support from the National Natural Science Foundation of China through Grant No. 11874083. 

\bibliography{prl_packet_beam_soi_v33_refs}

\end{document}


\title{Supplemental Material for\\
``Spectral Reduction Selects the Observable Imprint of Dirac Quantum Geometry''}

\author{Zhongze Guo}
\email{guozhongze007@gmail.com}
\affiliation{Department of Physics and Institute of Theoretical Physics, University of Science and Technology Beijing, Beijing 100083, China}

\author{Bei Xu}
\email{xubei0903@163.com}
\affiliation{Institute for Advanced Study, Tsinghua University, Beijing 100084, China}	

\author{Qiang Gu}
\email[Corresponding author: ] {qgu@ustb.edu.cn}
\affiliation{Department of Physics and Institute of Theoretical Physics, University of Science and Technology Beijing, Beijing 100083, China}

\maketitle

This Supplemental Material gives the technical derivations underlying the
Hermitian selection rule, its velocity-current corollary, its gauge
properties, the quantum-metric bound, and the exact Dirac realization
discussed in the main text. We use the same notation as in the main text. In particular, $P(\p)$ denotes the isolated internal projector, $Q=1-P$, and the free positive-energy Dirac frame is
\begin{equation}
U(\p)=\sqrt{\frac{E+mc^2}{2E}}
\begin{pmatrix}
\mathbf{1}_2\\[2mm]
\dfrac{c\sig\cdot\p}{E+mc^2}
\end{pmatrix},
\qquad E=\sqrt{m^2c^4+c^2p^2}.
\label{S:U}
\end{equation}

\section{Dirac preliminaries and inter-subspace identities}
\label{S:DiracPrelim}

The frame in Eq.~\eqref{S:U} obeys
\begin{equation}
U^\dagger U=\mathbf{1}_2,
\qquad
P=UU^\dagger,
\qquad
Q=1-P,
\end{equation}
and the projected current identity
\begin{equation}
{
U^\dagger\alpha_iU=\frac{cp_i}{E}\,\mathbf{1}_2.}
\label{S:projcurrent}
\end{equation}
In particular,
\begin{equation}
{
U^\dagger(p_x\alpha_y-p_y\alpha_x)U=0.}
\label{S:azcurrentzero}
\end{equation}
This identity removes radial-envelope derivatives from the axial current
moment and, below, implies the vanishing normal intraband symbol for the
Dirac magnetic moment.

A second identity exposes directly why the Dirac current vertex couples to
the inter-subspace geometry. With
\begin{equation}
H_D=c\bm\alpha\cdot\p+\beta mc^2,
\qquad H_DU=EU,
\end{equation}
differentiation followed by projection with $Q$ gives
\begin{equation}
Q(\partial_iH_D)U+QH_D\partial_iU=E\,Q\partial_iU.
\end{equation}
Since the complementary Dirac subspace has energy $-E$,
$QH_D=-EQ$, and $\partial_iH_D=c\alpha_i$. Hence
\begin{equation}
{
Qc\alpha_iU=2E\,Q\partial_iU.}
\label{S:interbandDirac}
\end{equation}
The projector $Q$ here does not represent a prepared negative-energy
particle; it records the component of the momentum derivative of the
positive-energy eigenspace outside that eigenspace.

\section{Exact coarea reduction and the Hermitian selection rule}
\label{S:coarea}

\subsection{Local spectral coordinates and the coarea fiber}
Let $\varepsilon(\p)$ be a scalar dispersion and consider a regular patch of
the energy surface
\begin{equation}
\Sigma_E=\{\p:\varepsilon(\p)=E\}
\end{equation}
for which
\begin{equation}
v_n\equiv\partial_{p_n}\varepsilon\neq0.
\end{equation}
The implicit-function theorem then permits the local coordinates
$(p_a,E)$, with $a=1,\ldots,d-1$, defined by
$p_n=p_n(p_a,E)$. Writing $v_i=\partial_{p_i}\varepsilon$, the derivatives at
fixed energy are
\begin{equation}
D_a\equiv\left.\partial_{p_a}\right|_E
=\partial_{p_a}-\frac{v_a}{v_n}\partial_{p_n},
\qquad
\partial_E=\frac{1}{v_n}\partial_{p_n}.
\label{S:Da}
\end{equation}
Consequently,
\begin{equation}
\partial_{p_a}=D_a+v_a\partial_E.
\label{S:derdecomp}
\end{equation}
Here and below, ``normal'' refers to the transverse coordinate direction
eliminated in this local spectral chart; it need not coincide with the
orthogonal Euclidean normal to $\Sigma_E$.

The coordinate Jacobian is
\begin{equation}
\dd^d p=\frac{\dd E\,\dd^{d-1}p_a}{|v_n|}.
\label{S:coareameasure}
\end{equation}
Thus the ambient Hilbert space is locally represented as a direct integral of
fixed-energy fibers,
\begin{equation}
L^2(\dd^d p)\simeq
\int^{\oplus}\dd E\,\mathcal H_E,
\qquad
\mathcal H_E=L^2(\dd^{d-1}p_a),
\label{S:directintegral}
\end{equation}
after the half-density rescaling
\begin{equation}
(\mathcal U_c\Psi)(E,p_a)
=|v_n|^{-1/2}\Psi\bigl(p_a,p_n(E,p_a)\bigr).
\label{S:halfden}
\end{equation}
Equation~\eqref{S:directintegral}, rather than a unitary equivalence between
the full ambient space and one individual shell, is the precise Hilbert-space
meaning of the coarea decomposition.

\subsection{Transformation of a symmetrically ordered first-order operator}
Consider in this chart a Hermitian first-order differential operator
\begin{equation}
\hat O_3=W+i\hbar\sum_a
\left(B_a\partial_{p_a}+\frac12\partial_{p_a}B_a\right),
\qquad W=W^\dagger,\quad B_a=B_a^\dagger,
\label{S:O3}
\end{equation}
where the derivative principal part is written along the coordinates retained
on the shell. A useful property of half-densities is that the symmetric
first-order operator associated with a vector field keeps the same form under
a coordinate change. Applying Eqs.~\eqref{S:halfden} and
\eqref{S:derdecomp} gives
\begin{align}
\mathcal U_c\hat O_3\mathcal U_c^{-1}
={}&W+i\hbar\sum_a
\left(B_aD_a+\frac12D_aB_a\right)
\nonumber\\
&+i\hbar\left(B_E\partial_E+\frac12\partial_EB_E\right),
\label{S:Osplit}
\end{align}
with
\begin{equation}
B_E\equiv\sum_a v_aB_a=v_nB_N,
\qquad
B_N\equiv\sum_a\frac{v_a}{v_n}B_a.
\label{S:BE}
\end{equation}
The first line of Eq.~\eqref{S:Osplit} acts tangentially within each
fixed-energy fiber, while the second line differentiates between neighboring
fibers. In the main text, \emph{spectral reduction} is the exact passage to a
fixed-$E$ coarea fiber while retaining the shell-tangent Hermitian operator,
\begin{equation}
\hat O_E=W+i\hbar\sum_a
\left(B_aD_a+\frac12D_aB_a\right).
\label{S:OE}
\end{equation}
No paraxial or near-shell expansion is involved in this definition.

For completeness, Eq.~\eqref{S:Osplit} can also be verified directly. Before
the half-density rescaling, the Jacobian $|v_n|^{-1}$ generates derivatives
of $\ln|v_n|$. Conjugation by $|v_n|^{-1/2}$ contributes precisely one half of
the corresponding divergence, so those Jacobian terms combine with the
original symmetric-ordering term into
$\frac12D_aB_a+\frac12\partial_EB_E$. This is the origin of the standard
half-density covariance of Eq.~\eqref{S:Osplit}.

\subsection{Projection to an isolated internal subspace}
Let $P(E,p_a)$ be an isolated internal projector. The difference between the
reduced and ambient operators is the negative of the energy-normal generator,
\begin{equation}
\delta\hat O\equiv \hat O_E-\hat O_3
=-i\hbar\left(B_E\partial_E+\frac12\partial_EB_E\right),
\label{S:dOraw}
\end{equation}
where Eq.~\eqref{S:dOraw} is understood after the coarea transform. Consider a
state in the isolated subspace, $\Psi=P\Psi$. Assume that the normal intraband symbol vanishes locally in a
neighborhood of the spectral shell,
\begin{equation}
PB_EP=0,
\label{S:PBEP0}
\end{equation}
which is equivalent on the regular chart to $PB_NP=0$. The neighborhood
condition is what permits the normal derivative of this identity used below.
Then
\begin{align}
PB_E\partial_E(P\Psi)
&=PB_EQ(\partial_EP)P\Psi
\equiv\mathcal C_E\Psi,
\label{S:Cfirst}
\\
\mathcal C_E&\equiv PB_EQ(\partial_EP)P.
\label{S:CE}
\end{align}
The derivative of the envelope inside the band has disappeared because its
coefficient is $PB_EP$. This is the precise sense in which condition
\eqref{S:PBEP0} is sufficient for extension independence. It is not claimed
to be necessary: if $PB_NP\neq0$, special cancellations may still occur for
particular observables or state families.

The factor $1/2$ in the final selection rule follows directly from
Hermiticity. Differentiating Eq.~\eqref{S:PBEP0} with respect to $E$ and
sandwiching by $P$ gives
\begin{equation}
P(\partial_EB_E)P
=-\mathcal C_E-\mathcal C_E^\dagger,
\label{S:dBE}
\end{equation}
because $P(\partial_EP)P=0$. Hence
\begin{align}
P\left(B_E\partial_E+\frac12\partial_EB_E\right)P
&=\mathcal C_E-\frac12
(\mathcal C_E+\mathcal C_E^\dagger)
\nonumber\\
&=\frac12(\mathcal C_E-\mathcal C_E^\dagger).
\label{S:antiherm}
\end{align}
Substituting this into Eq.~\eqref{S:dOraw} yields
\begin{equation}
{
\delta O_{\rm geom}
=-\frac{i\hbar}{2}
\left(\mathcal C_E-\mathcal C_E^\dagger\right).}
\label{S:selectionE}
\end{equation}
Finally, using Eq.~\eqref{S:BE} and
$\partial_E=v_n^{-1}\partial_{p_n}$,
\begin{equation}
\mathcal C_E
=PB_NQ(\partial_{p_n}P)P\equiv\mathcal C_N,
\label{S:CNC}
\end{equation}
which reproduces the Hermitian selection rule of the main text,
\begin{equation}
{
\delta O_{\rm geom}
=-\frac{i\hbar}{2}
\left(\mathcal C_N-\mathcal C_N^\dagger\right),
\qquad
\mathcal C_N=PB_NQ(\partial_{p_n}P)P.}
\label{S:selectionN}
\end{equation}
If $PB_NP\neq0$, the normal derivative of the intraband envelope can survive;
additional terms may then depend on how a shell state is continued away from
the shell. Equation~\eqref{S:selectionN} therefore gives a sufficient
extension-independent selection rule, not an if-and-only-if condition. The
Hermitian completion is essential: the unsymmetrized quantity
$-i\hbar\,U^\dagger B_NQ(\partial_{p_n}U)$ need not itself be Hermitian,
whereas Eq.~\eqref{S:selectionN} is Hermitian by construction and follows
from the symmetrically ordered ambient operator.

\subsection{Velocity-current magnetic moments as a general corollary}
\label{S:velocitycurrent}

The vanishing normal intraband symbol required by the selection rule is not
specific to the Dirac algebra. Consider a smooth translationally invariant
multicomponent Hamiltonian $H(\mathbf p)$ written in a momentum-independent
internal basis, with canonical position
\begin{equation}
	\hat r_i=i\hbar\partial_{p_i},
\end{equation}
and velocity and scalar-charge current operators
\begin{equation}
	\widehat{\mathcal V}_i=\partial_{p_i}H,
	\qquad
	\hat j_i=q\,\widehat{\mathcal V}_i .
	\label{S:velocitycurrent}
\end{equation}
Let $P=UU^\dagger$ project onto a smooth isolated band or an exactly
degenerate subspace with scalar dispersion,
\begin{equation}
	HP=\varepsilon P,
	\qquad
	v_i\equiv\partial_{p_i}\varepsilon,
	\qquad
	Q=1-P.
	\label{S:scalarband}
\end{equation}

The symmetrically ordered $z$ component of the first moment of this current is
\begin{equation}
	\hat M_z=
	\frac{q}{4}
	\left(
	\{\hat r_x,\widehat{\mathcal V}_y\}
	-
	\{\hat r_y,\widehat{\mathcal V}_x\}
	\right).
	\label{S:generalcurrentmoment}
\end{equation}
It has the first-order form used in the main text, with
\begin{equation}
	B_x=\frac{q}{2}\widehat{\mathcal V}_y,
	\qquad
	B_y=-\frac{q}{2}\widehat{\mathcal V}_x .
	\label{S:generalBxy}
\end{equation}

Differentiating $HP=\varepsilon P$ and projecting on both sides with $P$
gives the degenerate-subspace Hellmann--Feynman identity
\begin{equation}
	P(\partial_{p_i}H)P=v_iP.
	\label{S:HFdegenerate}
\end{equation}
For a regular fixed-energy chart in which $p_z$ is eliminated,
\begin{equation}
	B_E=v_xB_x+v_yB_y
	=
	\frac{q}{2}
	\left(
	v_x\widehat{\mathcal V}_y
	-
	v_y\widehat{\mathcal V}_x
	\right),
\end{equation}
and therefore
\begin{equation}
	PB_EP
	=
	\frac{q}{2}(v_xv_y-v_yv_x)P
	=0.
	\label{S:PBEPcurrent}
\end{equation}
Because Eq.~\eqref{S:HFdegenerate} holds throughout any neighborhood in which
the isolated scalar-dispersion subspace remains smooth, the neighborhood
condition required in the derivation of Eq.~\eqref{S:selectionN} is
automatically satisfied. On a regular chart with $v_z\neq0$,
$B_N=B_E/v_z$, so the extension-independent Hermitian selection rule applies
to this entire class of velocity-current moments.

The resulting geometry can be written in a form that also clarifies what is
special about the free Dirac problem. Differentiating
$HU=\varepsilon U$ and projecting into the complementary subspace gives
\begin{equation}
	Q\widehat{\mathcal V}_iU
	=
	(\varepsilon-H_Q)\,
	Q(\partial_{p_i}U),
	\qquad
	H_Q\equiv QHQ\big|_{\mathrm{Ran}\,Q}.
	\label{S:generalinterbandvelocity}
\end{equation}
Define the gap-weighted inter-subspace tensor
\begin{equation}
	\Xi_{ij}
	=
	(Q\partial_{p_i}U)^\dagger
	(\varepsilon-H_Q)
	(Q\partial_{p_j}U),
	\qquad
	\Xi_{ji}=\Xi_{ij}^\dagger,
	\label{S:Xi}
\end{equation}
and its Hermitian antisymmetric part
\begin{equation}
	\mathsf F^{\rm gap}_{ij}
	\equiv
	i(\Xi_{ij}-\Xi_{ji}).
	\label{S:Fgap}
\end{equation}
Using Eq.~\eqref{S:generalBxy} in the band representation of
Eq.~\eqref{S:selectionN} gives
\begin{equation}
	U^\dagger\mathcal C_NU
	=
	\frac{q}{2v_z}
	\left(
	v_x\Xi_{yz}-v_y\Xi_{xz}
	\right),
\end{equation}
and hence
\begin{equation}
	\delta\mathfrak M_z
	=
	-\frac{q\hbar}{4v_z}
	\left(
	v_x\mathsf F^{\rm gap}_{yz}
	-
	v_y\mathsf F^{\rm gap}_{xz}
	\right).
	\label{S:generalcurrentcorrection}
\end{equation}

In general, $\mathsf F^{\rm gap}_{ij}$ is not the non-Abelian Berry
curvature: different complementary bands carry different energy gaps and
cannot in general be factored into a common scalar. For the free Dirac
doublet, however, the complementary negative-energy subspace has
$H_Q=-E$ while $\varepsilon=E$, so
\begin{equation}
	\Xi_{ij}=2E\,\mathsf T_{ij},
	\qquad
	\mathsf F^{\rm gap}_{ij}=2E\,\mathsf F_{ij}.
	\label{S:Diracgapclosure}
\end{equation}
Together with $v_a/v_z=p_a/p_z$, Eq.~\eqref{S:generalcurrentcorrection}
reduces exactly to the Berry-curvature contraction for the Dirac magnetic
moment given in the main text. Thus the applicability of the Hermitian
selection rule is broader than the Dirac model, whereas the pure
Berry-curvature closure is a special consequence of its common interband gap.

This corollary assumes a momentum-independent microscopic internal basis,
canonical position, a scalar charge $q$ shared by the internal components, and
a scalar dispersion within the retained subspace. Momentum-dependent Bloch
bases, non-scalar charge vertices such as in Bogoliubov--de Gennes problems,
spin currents, and singular patches with $v_z=0$ require additional
structure and are not covered by the statement above.

\subsection{Two orders of reduction and a controlled packet-to-shell comparison}
It is useful to separate two operations: (i) compress the ambient operator to
a fixed-energy fiber only after projecting to the isolated subspace, and (ii)
first form the exact reduced shell operator and then project it to the same
subspace. Let
\begin{equation}
\widetilde O_3\equiv \mathcal U_c\hat O_3\mathcal U_c^{-1}
\end{equation}
denote the coarea-transformed ambient operator. Define the two band-space
fiber operators
\begin{equation}
\widehat A_E\equiv [U^\dagger \widetilde O_3 U]_E,
\qquad
\widehat R_E\equiv U_E^\dagger \hat O_E U_E.
\end{equation}
The Hermitian selection rule obtained above states precisely that
\begin{equation}
\widehat R_E-\widehat A_E
=\delta\mathfrak O_{\rm geom}(E),
\qquad
\delta\mathfrak O_{\rm geom}=U^\dagger\delta O_{\rm geom}U.
\label{S:AEminusRE}
\end{equation}
For a shell spinor $\eta$ in the common domain, the corresponding quadratic
forms are $\mathcal A_E[\eta]=\langle\eta,\widehat A_E\eta\rangle$ and
$\mathcal R_E[\eta]=\langle\eta,\widehat R_E\eta\rangle$. Thus the geometric
correction compares two different orders of reduction; it is not the ambiguity
of a single already-reduced decomposable operator.

To connect this fiber statement with normalized three-dimensional wave packets,
fix a compact set $K$ in a regular spectral chart and choose $\delta>0$ such
that
\begin{equation}
|v_n(E,p_a)|\ge v_*>0
\qquad
\text{for }(E,p_a)\in[E_0-\delta,E_0+\delta]\times K.
\label{S:packetregularity}
\end{equation}
Let $\eta\in C_c^\infty(K;\mathbb C^r)$ satisfy
\begin{equation}
\int_K d^{d-1}p_a\,\eta^\dagger\eta=1.
\label{S:etanorm}
\end{equation}
Choose $h\in C_c^\infty(\mathbb R)$ with $\|h\|_2=1$ and
$\operatorname{supp}h\subset[-1,1]$, set
$h_\epsilon(u)=\epsilon^{-1/2}h(u/\epsilon)$, and take
$0<\epsilon<\delta$. Define the ambient packet family
\begin{equation}
\widetilde\Psi_\epsilon(\mathbf p)
=|v_n(\mathbf p)|^{1/2}
 h_\epsilon\bigl(\varepsilon(\mathbf p)-E_0\bigr)
 U(\mathbf p)\eta(p_a).
\label{S:packetfamily}
\end{equation}
By the coarea formula and Eq.~\eqref{S:etanorm},
$\|\widetilde\Psi_\epsilon\|_2=1$, and the corresponding fiber amplitude is
exactly $h_\epsilon(E-E_0)U_E\eta$. For any bounded fiberwise continuous band
operator $\widehat A_E$, dominated convergence gives
\begin{equation}
\int dE\,|h_\epsilon(E-E_0)|^2\,\mathcal A_E[\eta]
\longrightarrow
\mathcal A_{E_0}[\eta],
\qquad \epsilon\to0.
\end{equation}
For the Dirac magnetic moment, the quadratic form on the fixed compact momentum
support is likewise continuous and bounded in $E$; therefore the same
argument gives
\begin{equation}
M_z^{(E_0)}[\eta]-\lim_{\epsilon\to0}M_z^{(3)}[\widetilde\Psi_\epsilon]
=\langle \eta,\delta\mathfrak M_z(E_0)\eta\rangle.
\label{S:packetlimit}
\end{equation}
This identifies the packet--beam difference as the shell correction at matched
transverse spectral probability. It does not claim equality of the resulting
coordinate-space fields, which differ by the fixed-energy Jacobian weight.

\section{Gauge covariance, spectral reparametrization, and chart locality}
\label{S:gauge}

Under a smooth change of local frame within the isolated rank-$r$ subspace,
\begin{equation}
U(\p)\longrightarrow U'(\p)=U(\p)G(\p),
\qquad G(\p)\in U(r),
\label{S:gaugeU}
\end{equation}
the projectors $P=UU^\dagger$ and $Q=1-P$ are unchanged. Therefore
Eq.~\eqref{S:selectionN}, written as an operator in the ambient internal
space, is manifestly frame independent. In an $r\times r$ internal matrix
representation,
\begin{equation}
U^\dagger\delta O_{\rm geom}U
\longrightarrow
G^\dagger\left(U^\dagger\delta O_{\rm geom}U\right)G,
\label{S:gaugecov}
\end{equation}
so expectation values are gauge invariant when the internal spinor transforms
as $\chi\to G^\dagger\chi$.

The same statement holds for the non-Abelian quantum geometric tensor
\begin{equation}
\mathsf T_{ij}=(Q\partial_iU)^\dagger(Q\partial_jU)
=\mathsf g_{ij}-\frac{i}{2}\mathsf F_{ij},
\label{S:QGT}
\end{equation}
because
\begin{equation}
Q\partial_i(UG)=(Q\partial_iU)G,
\end{equation}
and therefore
\begin{equation}
\mathsf T_{ij}\to G^\dagger\mathsf T_{ij}G,
\quad
\mathsf g_{ij}\to G^\dagger\mathsf g_{ij}G,
\quad
\mathsf F_{ij}\to G^\dagger\mathsf F_{ij}G.
\label{S:QGTcov}
\end{equation}

There is also a useful invariance under reparametrization of the same spectral
surface. If $E'=f(E)$ with $f'(E)\neq0$, then
\begin{equation}
B_{E'}=f'(E)B_E,
\qquad
\partial_{E'}=\frac{1}{f'(E)}\partial_E,
\end{equation}
so
\begin{equation}
PB_{E'}Q(\partial_{E'}P)P
=PB_EQ(\partial_EP)P.
\label{S:reparam}
\end{equation}
Thus the geometric product entering the selection rule does not depend on how
the same energy surface is labeled.

The construction is nevertheless local in a regular spectral chart. No claim
is made that an arbitrary change of the transverse lift away from the shell
leaves the separated normal generator unchanged. For the Dirac application in
the main text, regularity requires $p_z\neq0$. The physical
forward-propagating branch further restricts to $p_z>0$, with $(p_x,p_y)$
retained and $p_z=p_z(p_x,p_y;E)$ eliminated.

\section{Quantum-metric bound and equality condition}
\label{S:metricbound}

For a normalized internal spinor $\chi$, define the two inter-subspace vectors
\begin{equation}
|a_\chi\rangle=QB_NU|\chi\rangle,
\qquad
|b_\chi\rangle=Q(\partial_{p_n}U)|\chi\rangle.
\label{S:ab}
\end{equation}
Using $(\partial_{p_n}P)U=Q(\partial_{p_n}U)$, the expectation value of
Eq.~\eqref{S:selectionN} is
\begin{equation}
\chi^\dagger\delta\mathsf O_{\rm geom}\chi
=\hbar\,\operatorname{Im}\langle a_\chi|b_\chi\rangle.
\label{S:Imab}
\end{equation}
Therefore
\begin{align}
|\chi^\dagger\delta\mathsf O_{\rm geom}\chi|^2
&\le\hbar^2|\langle a_\chi|b_\chi\rangle|^2
\nonumber\\
&\le\hbar^2
\langle a_\chi|a_\chi\rangle
\langle b_\chi|b_\chi\rangle
\nonumber\\
&=\hbar^2
(\chi^\dagger\mathsf V_N\chi)
(\chi^\dagger\mathsf g_{nn}\chi),
\label{S:bound}
\end{align}
with
\begin{equation}
\mathsf V_N=U^\dagger B_NQB_NU,
\qquad
\mathsf g_{nn}=(Q\partial_{p_n}U)^\dagger
(Q\partial_{p_n}U).
\label{S:VNgnn}
\end{equation}
This is the quantum-metric bound stated in the main text. Equality requires both saturation of
Cauchy--Schwarz and a purely imaginary overlap; equivalently, the two
inter-subspace vectors must be locally proportional with a relative phase
$\pm\pi/2$ for the maximizing polarization.

\section{Quantum geometry of the free positive-energy Dirac doublet}
\label{S:DiracQGT}

Direct differentiation of Eq.~\eqref{S:U} gives a particularly simple
non-Abelian quantum geometric tensor. Its symmetric part is proportional to
the identity,
\begin{equation}
{
\mathsf g_{ij}
=\frac{c^2}{4E^2}
\left(\delta_{ij}-\frac{c^2p_ip_j}{E^2}\right)\mathbf{1}_2.}
\label{S:gij}
\end{equation}
The curvature can be written as
\begin{equation}
\mathsf F_{ij}=\epsilon_{ijk}\Omega_k,
\label{S:Fomega}
\end{equation}
with the matrix-valued momentum-space Berry curvature
\begin{equation}
{
\boldsymbol\Omega
=-\frac{c^4}{2E^3}
\left[
 m\sig+
 \frac{\p(\p\cdot\sig)}{E+mc^2}
\right].}
\label{S:Omega}
\end{equation}
Equations~\eqref{S:gij} and \eqref{S:Omega} follow from
$\mathsf T_{ij}=(Q\partial_iU)^\dagger(Q\partial_jU)$ with the convention
$\mathsf T_{ij}=\mathsf g_{ij}-i\mathsf F_{ij}/2$.

Two components used in the main text follow immediately. First,
\begin{equation}
\mathsf g_{zz}
=\frac{c^2}{4E^2}
\left(1-\frac{c^2p_z^2}{E^2}\right)\mathbf{1}_2.
\label{S:gzz}
\end{equation}
Second, for the angular derivative at fixed $(p_\perp,p_z)$,
$\partial_\varphi\p=(-p_y,p_x,0)$, and Eq.~\eqref{S:gij} gives
\begin{equation}
\mathsf g_{\varphi\varphi}
=g_{\varphi\varphi}\1_2,
\qquad
g_{\varphi\varphi}=\frac{c^2p_\perp^2}{4E^2}.
\label{S:gphiphi}
\end{equation}
Hence
\begin{equation}
{
\Delta=\frac{c^2p_\perp^2}{E(E+mc^2)}
=\frac{4E}{E+mc^2}\,g_{\varphi\varphi},}
\label{S:Deltag}
\end{equation}
which establishes the exact angular quantum-metric interpretation of the
standard Dirac spin--orbit scale $\Delta$.

\subsection{Doublet metric versus a selected-state line metric}
The metric in Eqs.~\eqref{S:gij}--\eqref{S:Deltag} is the inter-subspace
metric of the full rank-two positive-energy projector. It should not be
confused with the Abelian line-bundle metric obtained after choosing one
particular normalized state $u_s=U\chi_s$ inside that doublet. For a
longitudinal spinor, $\sigma_z\chi_s=2s\chi_s$, the latter is
\begin{equation}
 g_{\varphi\varphi}^{\rm line}
 =\langle\partial_\varphi u_s|
 (1-|u_s\rangle\langle u_s|)|\partial_\varphi u_s\rangle.
\end{equation}
A direct evaluation gives
\begin{equation}
{
 g_{\varphi\varphi}^{\rm line}
 =\frac{\Delta}{2}-\frac{\Delta^2}{4}.}
\label{S:linemetric}
\end{equation}
The difference from Eq.~\eqref{S:gphiphi} is expected: the line-bundle
projection counts motion both into the negative-energy sector and into the
orthogonal state within the positive-energy doublet, whereas the non-Abelian
metric used in the Letter measures only motion out of the entire
positive-energy subspace. This distinction removes an otherwise easy source
of factor-of-two or convention confusion.

\subsection{Pullback of the metric to the fixed-energy shell}
Writing the shell derivatives as
\begin{equation}
D_a=t_a^{\ i}\partial_i,
\qquad
t_a^{\ i}=\delta_a^{\ i}-\frac{v_a}{v_n}\delta_n^{\ i},
\end{equation}
the quantum geometric tensor restricted to the shell is simply the tensor
pullback
\begin{equation}
{
\mathsf T^{\Sigma}_{ab}=t_a^{\ i}t_b^{\ j}\mathsf T_{ij},
\qquad
\mathsf g^{\Sigma}_{ab}=t_a^{\ i}t_b^{\ j}\mathsf g_{ij}.}
\label{S:pullback}
\end{equation}
Thus spectral reduction does not erase the local projector geometry. For the
free Dirac shell with $n=z$ and $v_a/v_z=p_a/p_z$,
Eq.~\eqref{S:gij} yields
\begin{equation}
{
\mathsf g^{(E)}_{ab}
=\frac{c^2}{4E^2}
\left(\delta_{ab}+\frac{p_ap_b}{p_z^2}\right)\1_2,
\qquad a,b=x,y.}
\label{S:pullbackDirac}
\end{equation}
The metric therefore behaves in the standard geometric way under the shell
restriction; the special effect discussed in the Letter comes from the
change in the differential realization of the position-weighted magnetic
observable.

\section{Dirac magnetic moment: selection rule, curvature contraction, and metric saturation}
\label{S:DiracM}

For
\begin{equation}
\hat M_z^{\rm amb}
=\frac{qi\hbar c}{2}
(\alpha_y\partial_{p_x}-\alpha_x\partial_{p_y}),
\end{equation}
the first-order coefficients are
\begin{equation}
B_x=\frac{qc}{2}\alpha_y,
\qquad
B_y=-\frac{qc}{2}\alpha_x.
\end{equation}
On the fixed-energy Dirac shell,
$v_i=c^2p_i/E$, so for the chart eliminating $p_z$,
\begin{equation}
B_N=\frac{qc}{2p_z}
(p_x\alpha_y-p_y\alpha_x).
\label{S:BNdirac}
\end{equation}
Using
$U^\dagger\boldsymbol\alpha U=(c\p/E)\1_2$ gives
\begin{equation}
U^\dagger B_NU=0,
\qquad PB_NP=0,
\label{S:PBdirac}
\end{equation}
pointwise throughout the regular chart, so in particular in a neighborhood
of any fixed-energy shell under consideration. Thus the
extension-independent rule Eq.~\eqref{S:selectionN} applies exactly.

Equation~\eqref{S:interbandDirac} makes the geometric structure explicit.
Defining
$\Gamma_N=U^\dagger B_NQ(\partial_{p_z}U)$, one obtains
\begin{equation}
\Gamma_N
=\frac{qE}{p_z}
\left(p_x\mathsf T_{yz}-p_y\mathsf T_{xz}\right).
\label{S:GammaQGT}
\end{equation}
Taking the anti-Hermitian part and using
$\mathsf T_{ij}=\mathsf g_{ij}-i\mathsf F_{ij}/2$ immediately gives
\begin{equation}
{
\delta\mathsf M_z
=-\frac{q\hbar E}{2p_z}
\left(p_x\mathsf F_{yz}-p_y\mathsf F_{xz}\right).}
\label{S:dMcurvStructural}
\end{equation}
Thus the curvature contraction follows directly from the Dirac current
vertex and the inter-subspace derivative identity, rather than only from an
explicit substitution of the free-particle curvature.

For reference, the complete ambient spinor-derivative contribution is
\begin{equation}
\frac{qi\hbar c}{2}
U^\dagger(\alpha_y\partial_{p_x}-\alpha_x\partial_{p_y})U
=\mu_E\sigma_z+\mathcal G_z(\p),
\label{S:fullframederivative}
\end{equation}
where the residual geometry-sensitive kernel is
\begin{equation}
\mathcal G_z(\p)
=-\frac{q\hbar c^4}{4E^2(E+mc^2)}
\left[p_\perp^2\sigma_z-p_z\pperp\cdot\sig_\perp\right].
\label{S:Gz}
\end{equation}
In particular, at $\p=0$ the residual kernel vanishes while the complete
spinor derivative reduces to $\mu_E\sigma_z$, providing a useful consistency
check on the separation. A direct evaluation of the shell-induced difference
then gives
\begin{equation}
\delta\mathsf M_z
=\frac{q\hbar c^2}{4Ep_z}
\pperp\cdot\sig_\perp-\mathcal G_z(\p).
\label{S:dMdirect}
\end{equation}
The same result follows directly from the curvature. Equations
\eqref{S:Fomega}--\eqref{S:Omega} imply
\begin{align}
p_x\mathsf F_{yz}-p_y\mathsf F_{xz}
={}&-\frac{c^4}{2E^3}
\Bigg[
\left(m+\frac{p_\perp^2}{E+mc^2}\right)
\pperp\cdot\sig_\perp
\nonumber\\
&\hspace{24mm}
+\frac{p_\perp^2p_z}{E+mc^2}\sigma_z
\Bigg].
\label{S:Fcontraction}
\end{align}
Multiplying Eq.~\eqref{S:Fcontraction} by
$-q\hbar E/(2p_z)$ and using
$E^2=m^2c^4+c^2p^2$ reproduces Eq.~\eqref{S:dMdirect} and independently
checks Eq.~\eqref{S:dMcurvStructural}. The curvature relation is therefore an
exact operator identity within the positive-energy doublet.

The metric bound is also exactly saturable. Since
$PB_NP=0$, one may use $QB_NU=B_NU$, and the Dirac algebra gives
\begin{equation}
\mathsf V_N
=U^\dagger B_NQB_NU
=\frac{q^2c^2p_\perp^2}{4p_z^2}\1_2.
\label{S:VNdirac}
\end{equation}
Combining Eqs.~\eqref{S:gzz} and \eqref{S:VNdirac},
\begin{equation}
\hbar^2\mathsf V_N\mathsf g_{zz}
=\frac{\mu_E^2}{4}\tan^2\theta
\left(\sin^2\theta+\frac{\cos^2\theta}{\gamma^2}\right)\1_2,
\label{S:boundDirac}
\end{equation}
where $p_\perp=p\sin\theta$, $p_z=p\cos\theta$, and
$\gamma=E/(mc^2)$. On the other hand,
\begin{equation}
\delta\mathsf M_z=A_r\sigma_r+A_z\sigma_z,
\label{S:dMcyl}
\end{equation}
with
\begin{align}
A_r&=\frac{\mu_E}{2}\tan\theta
\left(\sin^2\theta+\frac{\cos^2\theta}{\gamma}\right),
\nonumber\\
A_z&=\frac{\mu_E}{2}
\left(1-\frac1\gamma\right)\sin^2\theta,
\end{align}
and a direct algebraic reduction yields
\begin{equation}
{
(\delta\mathsf M_z)^2
=(A_r^2+A_z^2)\mathbf{1}_2
=\hbar^2\mathsf V_N\mathsf g_{zz}.}
\label{S:saturation}
\end{equation}
Thus the two eigenpolarizations of $\delta\mathsf M_z$ attain the metric bound
pointwise in momentum. Pointwise saturation does not by itself imply that one
momentum-independent polarization, or a globally constrained smooth texture,
saturates an integrated wave-packet bound over the entire momentum support.

\section{Transverse-helicity dark sectors and coherence-bright states}
\label{S:dark}

Introduce the cylindrical Pauli matrices
\begin{equation}
\sigma_r=\cos\varphi_p\,\sigma_x+\sin\varphi_p\,\sigma_y,
\qquad
\sigma_\varphi=-\sin\varphi_p\,\sigma_x+\cos\varphi_p\,\sigma_y.
\end{equation}
The conserved transverse-helicity operator
$\hat K=\beta(\bm\Sigma\times\p)_z$ projects onto the positive-energy doublet
as
\begin{equation}
\mathsf K\equiv U^\dagger\frac{\hat K}{p_\perp}U=-\sigma_\varphi,
\qquad (p_\perp\neq0).
\label{S:K}
\end{equation}
Since Eq.~\eqref{S:dMcyl} contains only $\sigma_r$ and $\sigma_z$,
\begin{equation}
\{\delta\mathsf M_z,\mathsf K\}=0.
\end{equation}
For the projectors $\Pi_\pm=(1\pm\mathsf K)/2$,
\begin{equation}
{\Pi_\pm\delta\mathsf M_z\Pi_\pm=0.}
\label{S:darkexact}
\end{equation}
Thus a density matrix diagonal in the $K$ resolution has zero shell-induced
magnetic response; inter-branch coherence is therefore necessary, though not
sufficient by itself, for a nonzero response.

For the family
\begin{equation}
\eta_{\ell,\delta}(\varphi_p)
=\frac{e^{i\ell\varphi_p}}{\sqrt2}
\begin{pmatrix}1\\ e^{i(\varphi_p+\delta)}\end{pmatrix},
\label{S:etad}
\end{equation}
one has
\begin{equation}
J_z\eta_{\ell,\delta}
=\hbar\left(\ell+\frac12\right)\eta_{\ell,\delta},
\end{equation}
and
\begin{equation}
\langle\sigma_r\rangle=\cos\delta,
\qquad
\langle\sigma_\varphi\rangle=\sin\delta,
\qquad
\langle\sigma_z\rangle=0.
\end{equation}
Therefore
\begin{equation}
\delta M_z(\delta)=A_r\cos\delta.
\label{S:dMdelta}
\end{equation}
The states $\delta=\pm\pi/2$ are the two definite-$K$ branches and are exactly
dark, while the radial texture $\delta=0$ is an equal-weight coherent
superposition and is bright. At fixed finite $\gamma$,
\begin{equation}
\delta M_z^{\rm rad}=A_r
=\frac{\mu_E}{2\gamma}\theta+O(\theta^3),
\end{equation}
whereas a longitudinal polarization gives
$\delta M_z=\mu_Es\Delta=O(\theta^2)$.

\section{Fixed-$p_z$ delocalized control}
\label{S:fixedpz}

The disappearance of the longitudinal $s\Delta$ magnetic term in a
monochromatic beam is not caused merely by delocalization along $z$. To see
this independently, consider a transversely normalizable state with a sharp
$p_z$ but a spread in $p_\perp$,
\begin{equation}
\Phi_{p_z}(\pperp,t)
=f(p_\perp)e^{i\ell\varphi_p}U(\p)\eta_s
 e^{-iE(\p)t/\hbar}.
\label{S:fixedpzstate}
\end{equation}
It is completely delocalized in $z$ but is not monochromatic, since $E$
varies across the transverse momentum distribution. Its transverse current
moment is evaluated with the ordinary ambient derivatives
$\partial_{p_x},\partial_{p_y}$. Derivatives of the radial envelope and of the
time-dependent phase are proportional to the same azimuthal current vertex
that vanishes by Eq.~\eqref{S:azcurrentzero}. The remaining vortex-phase and
frame-derivative terms are therefore exactly those of the ambient
calculation, giving
\begin{equation}
{
M_z^{(p_z)}
=\left\langle
\mu_E\bigl(\ell+2s-s\Delta\bigr)
\right\rangle_f.}
\label{S:fixedpzresult}
\end{equation}
Thus longitudinal delocalization alone does not remove the $s\Delta$
magnetic contribution; the distinction arises when the state is restricted
to a fixed-energy shell and the transverse derivatives become shell tangent.

\section{Stationary Gordon-current check and the origin of the packet--beam difference}
\label{S:Gordon}

A direct coordinate-space check verifies the fixed-energy magnetic moment and
clarifies the additional term generated when the longitudinal phase varies
across the energy shell. Let
\begin{equation}
\Psi(\mathbf r,t)=e^{-iEt/\hbar}\Psi(\mathbf r)
\end{equation}
be a free stationary Dirac solution. Using
$H\Psi=E\Psi$, the anticommutation relations of the Dirac matrices, and
$\alpha_i\alpha_j=\delta_{ij}+i\epsilon_{ijk}\Sigma_k$, one finds the exact
stationary Gordon identity
\begin{equation}
{
\mathbf j
=\frac{qc^2}{2E}
\left[
\Psi^\dagger\mathbf p\Psi
+(\mathbf p\Psi)^\dagger\Psi
+\hbar\nabla\times(\Psi^\dagger\bm\Sigma\Psi)
\right].}
\label{S:Gordonidentity}
\end{equation}
For a transversely normalizable beam define
\begin{equation}
\mathcal N_\perp=\int\dd^2r_\perp\,\Psi^\dagger\Psi,
\qquad
\mathcal M_z(z)=\frac12\int\dd^2r_\perp
\bigl[\mathbf r_\perp\times\mathbf j(\mathbf r_\perp,z)\bigr]_z.
\label{S:Mfirst}
\end{equation}
For the shell Fourier representation used below, Parseval's identity makes
$\mathcal N_\perp$ independent of $z$. Assuming the transverse boundary
terms vanish, Eq.~\eqref{S:Gordonidentity} gives
\begin{equation}
{
\mathcal M_z
=\mu_E\left[
\frac{\mathcal L_z}{\hbar}
+\frac{2\mathcal S_z}{\hbar}
+\frac12\partial_z\mathcal R_\perp
\right],}
\label{S:Gordonmoment}
\end{equation}
where
\begin{align}
\mathcal L_z&=\int\dd^2r_\perp\,\Psi^\dagger L_z\Psi,
\\
\mathcal S_z&=\frac{\hbar}{2}\int\dd^2r_\perp\,
\Psi^\dagger\Sigma_z\Psi,
\\
\mathcal R_\perp&=\int\dd^2r_\perp\,
\mathbf r_\perp\cdot
\Psi^\dagger\bm\Sigma_\perp\Psi.
\label{S:Rperp}
\end{align}
For a fixed-energy shell this last term is generally nonzero because
$p_z$ varies with $\pperp$; it is the coordinate-space counterpart of the
shell-tangent derivative correction.

To evaluate it, write the fixed-energy state as
\begin{equation}
\Psi(\mathbf r_\perp,z)
=\int\frac{\dd^2p_\perp}{(2\pi\hbar)^2}
\,\Phi(\pperp)
\exp\left[\frac{i}{\hbar}
(\pperp\cdot\mathbf r_\perp+p_z(\pperp)z)\right].
\label{S:shellFT}
\end{equation}
Using $\mathbf r_\perp\leftrightarrow i\hbar\nabla_{p_\perp}$ in the
transverse Fourier transform, differentiation with respect to $z$ gives
\begin{equation}
\partial_z\mathcal R_\perp
=-\sum_{a=x,y}
\int\frac{\dd^2p_\perp}{(2\pi\hbar)^2}\,
\Phi^\dagger\Sigma_a\Phi\,\partial_{p_a}p_z.
\label{S:dRgeneral}
\end{equation}
Introduce
\begin{equation}
\mathcal N_\perp\equiv\int\frac{\dd^2p_\perp}{(2\pi\hbar)^2}\,\Phi^\dagger\Phi,
\qquad
\langle A\rangle_\Phi\equiv
\frac{1}{\mathcal N_\perp}\int\frac{\dd^2p_\perp}{(2\pi\hbar)^2}\,\Phi^\dagger A\Phi.
\label{S:Nperpdef}
\end{equation}
On the free fixed-energy shell,
$\partial_{p_a}p_z=-p_a/p_z$, so
\begin{equation}
{
\frac{1}{2\mathcal N_\perp}\partial_z\mathcal R_\perp
=\frac12\left\langle
\frac{\pperp\cdot\bm\Sigma_\perp}{p_z}
\right\rangle_\Phi.}
\label{S:dRshell}
\end{equation}
This is precisely the contribution missed if one treats the longitudinal
phase as independent of transverse momentum.

The positive-energy projection of the Dirac spin operator is
\begin{equation}
U^\dagger\Sigma_iU
=\frac{mc^2}{E}\sigma_i
+\frac{c^2p_i(\p\cdot\sig)}{E(E+mc^2)}.
\label{S:SigmaProj}
\end{equation}
For longitudinal polarization, $\sigma_z\eta_s=2s\eta_s$, the transverse
components are
\begin{equation}
\eta_s^\dagger U^\dagger\Sigma_aU\eta_s
=\frac{2s c^2p_ap_z}{E(E+mc^2)},
\qquad a=x,y.
\label{S:SigmaperpLong}
\end{equation}
For a smooth transversely normalizable annular spectrum, define
\begin{equation}
\overline\Delta_\Phi\equiv\langle\Delta(\mathbf p_\perp)\rangle_\Phi.
\label{S:Deltabar}
\end{equation}
Equations~\eqref{S:dRshell} and \eqref{S:SigmaperpLong} then give
\begin{equation}
\frac{1}{2\mathcal N_\perp}\partial_z\mathcal R_\perp
=s\overline\Delta_\Phi.
\label{S:textureDelta}
\end{equation}
At the same time,
\begin{equation}
\frac{\mathcal S_z}{\hbar\mathcal N_\perp}
=s(1-\overline\Delta_\Phi),
\qquad
\frac{\mathcal L_z}{\hbar\mathcal N_\perp}
=\ell+s\overline\Delta_\Phi.
\label{S:SL}
\end{equation}
Hence Eq.~\eqref{S:Gordonmoment} yields, for any such fixed-energy annular
spectrum,
\begin{equation}
\frac{\mathcal M_z^{(E)}}{\mathcal N_\perp}
=\mu_E(\ell+2s).
\label{S:GordonFinal}
\end{equation}
A single-cone formula is understood as the zero-width limit of these smooth
annular spectra, with transverse integration and normalization performed before
the cone width is taken to zero; in that limit
$\overline\Delta_\Phi\to\Delta$.

Equation~\eqref{S:GordonFinal} is an independent coordinate-space check of
the fixed-energy result. A transversely normalizable fixed-$p_z$ packet with a
spread in $p_\perp$ is not, in general, a single-energy stationary state, so
the stationary identity above should not be used as a second derivation of its
magnetic moment. Its ambient result instead follows from the momentum-space
calculation in the main text. Together, the two calculations isolate the
fixed-energy spectral constraint---rather than longitudinal delocalization by
itself---as the operation responsible for the packet--beam difference.

\section{Special-limit consistency checks}
\label{S:limits}

The exact formulas admit several useful limiting checks.  They also clarify
which apparent singularities belong to the chosen spectral chart or internal
texture rather than to the positive-energy projector itself.

\paragraph{On-axis limit.}
At fixed $p_z\neq0$, $p_\perp\to0$ gives
\begin{equation}
\Delta\to0,\qquad \mathcal G_z\to0,\qquad
\delta\mathsf M_z\to0.
\end{equation}
The complete ambient spinor-derivative term nevertheless tends to
$\mu_E\sigma_z$, in agreement with Eq.~\eqref{S:fullframederivative}.  Thus
the residual kernel $\mathcal G_z$ is not the full spinor derivative.  The
normalized transverse-helicity operator $\hat K/p_\perp$ is undefined exactly
on the axis, so the $K$-resolved dark/bright statements are understood on the
finite-cone patch $p_\perp\neq0$ and may be approached continuously from it.

\paragraph{Paraxial limit at fixed finite $\gamma$.}
Using $\gamma=E/(mc^2)$ and $p_\perp=p\sin\theta$ gives the exact identity
\begin{equation}
\Delta=\left(1-\frac1\gamma\right)\sin^2\theta,
\qquad
A_z=\frac{\mu_E}{2}\Delta .
\end{equation}
Hence a longitudinal state with $\sigma_z\eta_s=2s\eta_s$ has
$2sA_z=\mu_Es\Delta=O(\theta^2)$.  The radial coefficient instead expands as
\begin{equation}
A_r=\frac{\mu_E}{2\gamma}\theta
+\frac{\mu_E}{2}\left(1-\frac{2}{3\gamma}\right)\theta^3
+O(\theta^5),
\end{equation}
which verifies the leading $O(\theta/\gamma)$ term quoted in the Letter.

\paragraph{Breakdown of the $p_z$ spectral chart.}
The formulas containing $1/p_z$ are derived in the chart in which $p_z$ is
eliminated by the fixed-energy constraint.  As $p_z\to0$,
$v_z=c^2p_z/E\to0$ and this chart ceases to be regular.  The resulting
coordinate divergence of $D_a$, $B_N$, or $\delta\mathsf M_z$ is therefore not
a physical singularity of the positive-energy subspace; the shell must be
covered by a different regular chart.

\paragraph{Nonrelativistic limit.}
For a smooth longitudinal polarization, $\gamma\to1^+$ implies
$\Delta\to0$, $A_z\to0$, and $\mathcal G_z\to0$, recovering the ordinary
Pauli magnetic moment.  If instead one approaches $\mathbf p=0$ at fixed cone
angle while keeping the radial spin texture, the formal coefficient tends to
\begin{equation}
A_r\longrightarrow \frac{\mu_{\rm NR}}{2}\tan\theta,
\qquad \mu_{\rm NR}=\frac{q\hbar}{2m}.
\end{equation}
This does not define a smooth rest-frame limit: the radial texture depends on
$\varphi_p$, and both $\varphi_p$ and the normalized $K$ resolution are
undefined at $\mathbf p=0$.  In the regular on-axis limit
$p_\perp\to0$ at fixed $p_z\neq0$, the correction instead vanishes as stated
above.

\paragraph{Ultrarelativistic and order-of-limits check.}
At fixed $\theta$ and $\gamma\to\infty$,
\begin{equation}
\Delta\to\sin^2\theta,\qquad
A_z\to\frac{\mu_E}{2}\sin^2\theta,\qquad
A_r\to\frac{\mu_E}{2}\tan\theta\,\sin^2\theta,
\end{equation}
while $\mu_E=q\hbar c^2/(2E)\to0$ in absolute magnitude.  If the
ultrarelativistic limit is taken before the paraxial expansion, the linear
term in $A_r$ is absent and $A_r\sim\mu_E\theta^3/2$.  The
$O(\theta/\gamma)$ statement in the Letter therefore refers explicitly to
$\theta\to0$ at fixed finite $\gamma$; for large $\gamma$, the linear term
dominates the cubic term in the regime $\gamma\theta^2\ll1$.

\paragraph{Angular-metric check.}
Finally,
\begin{equation}
 g_{\varphi\varphi}=\frac{c^2p_\perp^2}{4E^2},\qquad
 \Delta=\frac{4E}{E+mc^2}g_{\varphi\varphi}
\end{equation}
is dimensionally consistent and both sides vanish quadratically as
$p_\perp\to0$.  This also provides a direct convention check on the
rank-two positive-energy quantum metric used throughout the Letter.

\section{Assumptions, regularity, and convention ledger}
\label{S:ledger}

For clarity, the general selection rule used in the Letter assumes: (i) a
smooth isolated rank-$r$ projector $P(\p)$; (ii) scalar dispersion
$\varepsilon(\p)$ on that subspace; (iii) a regular local spectral chart with
$v_n\neq0$; (iv) a common smooth operator domain and boundary conditions
sufficient for the integrations by parts used above; (v) a symmetrically
ordered first-order observable with $W=W^\dagger$ and Hermitian principal
symbols $B_a=B_a^\dagger$; (vi) the coarea measure and half-density
identification of fixed-energy fibers; and (vii) $PB_NP=0$ locally in a
neighborhood of the shell for the extension-independent form
Eq.~\eqref{S:selectionN}. This last condition is sufficient, not asserted to
be necessary.

For the Dirac application, a regular spectral chart requires $p_z\neq0$;
the forward-propagating branch used in the main text has $p_z>0$. The
normalized transverse-helicity resolution $\hat K/p_\perp$ additionally
requires $p_\perp\neq0$. These are chart or resolution singularities, not
singularities of the positive-energy projector itself. The dark/bright
statements apply on the regular cylindrical patch $p_z\neq0$,
$p_\perp\neq0$, with the vortex states discussed in the Letter lying on its
forward subset $p_z>0$.

Our sign and normalization conventions are as follows. The symbol $q$ is the
particle charge (so $q=-e$ for an electron). The non-Abelian quantum geometric
tensor is
\begin{equation}
\mathsf T_{ij}=\mathsf g_{ij}-\frac{i}{2}\mathsf F_{ij},
\qquad
\mathsf F_{ij}=i(\mathsf T_{ij}-\mathsf T_{ji}).
\end{equation}
Changing the Berry-curvature sign convention changes corresponding signs in
the curvature representation but not the physical operator. The
$g_{\varphi\varphi}$ in Eq.~\eqref{S:Deltag} is the rank-two positive-energy
inter-subspace metric, not the selected-state line metric of
Eq.~\eqref{S:linemetric}. Finally, ``normal'' throughout denotes the
eliminated transverse coordinate direction $p_n$ of the chosen spectral
chart, not necessarily the orthogonal Euclidean normal to the energy surface.